Atacama
Large
Millimeter /
submillimeter
Array

**ALMA Memo 618**

# The ObsMode 2020 Process

Satoko Takahashi (ObsMode2020 lead, JAO/NAOJ),
Edward B. Fomalont (NRAO/ JAO), Yoshiharu Asaki (NAOJ/ JAO), Geoff Crew (MIT Haystack Observatory), Lynn D. Matthews (MIT Haystack Observatory), Paulo Cortes (JAO/ NRAO), Baltasar Vila-Vilaro (JAO/ ESO), Tim Bastian (NRAO), Masumi Shimojo (NAOJ), Andy Biggs (ESO), Hugo Messias (JAO/ ESO), Antonio Hales (JAO/ NRAO), Eric Villard (ESO), and Elizabeth Humphreys (JAO/ ESO)

## Table of Contents

# Abstract

ObsMode is a yearly process which aims at preparing capabilities for future Observing cycles. The process has been running for a number of years tied to each ALMA observing cycle, with various leaderships. This document specifically summarizes the ObsMode2020 process (April-October 2020) with a new scheme, which is based on a proposal by E. Villard (the former head of APG/ JAO). Some improvements and the actual implementation were performed by S. Takahashi (the ObsMode lead/ JAO).

In the ObsMode2020 process, seven capabilities are identified as high priority items, for which it was originally aimed to be ready for Cycle 9. However, because of the observatory shutdown due to the covid-19 pandemic, we were not able to obtain any test data. This forced us to revise our timelines and delay the testing plan by one year. After the revision of the timelines, the ObsMode2020 process was adjusted for preparation of the new Cycle 8, which starts from October 2021. In reality, most of the capabilities became no-go for Cycle 8 due to the lack of test data. These have been carried over to the ObsMode2021 process, which aims to be ready for Cycle 9.

Our time in 2020 was spent carefully preparing for the commissioning test plans, revisiting previously obtained test data sets, and working on subsystem requirements and implementations. The ObsMode review process was implemented for the first time, and test plan documents and technical summary reports were reviewed by experts from outside of the commissioning team.

While no new data were obtained during the observatory shutdown, verifications using the existing data allowed us to offer the 7m-array polarization capability (in ACA standalone mode, single field rather than mosaicking) for Cycle 8 starting from October, 2021. In addition, subsystem readiness and policy-side preparations for the phased array observing mode were improved for Cycle 8. Other high priority items were decided to be carried over to ObsMode2021.

# 1. Introduction

## 1.1. Definition and purpose

ObsMode is a yearly process which aims at preparing capabilities for future ALMA Observing cycles. The process has been running for a number of years tied to each ALMA observing cycle, with various leaderships. In 2020, we decided to implement a new ObsMode process, which is better tied to the yearly ALMA Cycle as well as ALMA subsystem implementation timelines. This idea was originally proposed by E. Villard (the former head of APG/ JAO) and approved by the ALMA Deputy Director and the Head of DSO. The actual implementation was performed by S. Takahashi (the ObsMode lead/ JAO) with additional considerations, adjustments, and improvements. This ALMA memo summarizes the ObsMode2020 process run between April and October 2020.

## 1.2. Acronyms

| | |
|---|---|
| ACA | Atacama Compact Array |
| ALMA | Atacama Large Millimeter/submillimeter Array |
| APG | Array Performance Group |
| APP2 | ALMA Phasing Project 2 |
| ARC | ALMA Regional Center |
| ASAC | The ALMA Scientific Advisory Committee |
| B2B | Band-to-Band |
| DMG | Data Management Group |
| DRM | Data Reduction Manager |
| DSO | Department of Science Operation |
| EDM | (ALMA) Electronic Document Management (http://edm.alma.cl) |
| EHT | Event Horizon Telescope |
| EOC | Extension and Optimization of Capabilities |
| FDM | Frequency Division Mode |
| FWHM | Full width half maximum |
| GMVA | The Global mm-VLBI Array |
| ICT | Information and Communication Technology |
| ISOpT | Integrated Science Operation Team |
| JAO | Joint ALMA Observatory |
| MIT | Massachusetts Institute of Technology |
| NA | North America |
| NRAO | National Radio Astronomical Observatory |
| OT | Observing Tool |
| P2G | Phase2 Group |
| PG | Proposer's Guide |
| SSG | Subsystem Group |
| SSR | Science Software Requirements |
| SSS | Subsystem Scientist |
| TDM | Time Division Mode |
| THB | Technical Hand Book |
| VLBI | Very Long Baseline Interferometry |

# 2. Description

## 2.1. ObsMode2020 process

The ObsMode2020 process was originally started in order to focus on Cycle 9 capability testing and development. E. Villard (head of APG) presented the preliminary ObsMode2020 priorities and timeline in February 2020 to EOC colleagues (JAO and ARC colleagues who focus on technically-oriented work and who are involved in development work). This provided notification and a discussion opportunity across the ALMA project. Candidate capabilities/observing modes were then presented at the ASAC meeting in early March 2020 in order to be compared with the interests of the community, and they were properly prioritized according to community requests. The final priorities for Cycle 9 capabilities were officially decided after feedback from the ASAC members. The outcome consisted of seven capabilities. Please note that each observing capability is identified as an equally high priority item, led by an independent technical lead. Sometimes more than one observing mode was identified as being associated with a particular capability. In this case, the modes have a relative priority based on community interest.

## 2.2. ObsMode2020 Timelines

Starting from 2020, the ObsMode process aligns better with the ALMA ONLINE/OFFLINE subsystem release timelines, which are also tied to the ALMA Cycles. This facilitates better collaborative work between the ObsMode process and SSGs, particularly with the OT Phase I and Phase II requirement deadlines. For this, the ObsMode process changed to an October-October timeline (from an April/May-December timeline). Year 2020 was the first year for which we adopted the new timeline, hence the process was shorter than will take place in future. ObsMode2020 started in April 2020 and concluded in October 2020. The ObsMode annual time schedule that was shared in April 2020, is presented in Table 1.

**Table 1:** ObsMode2020 annual timeline presented on April 6, 2020 in the ObsMode-SSG kick off telecon, and later adjusted with the new situation in 2020. The schedule presented here is the timeline we followed for the ObsMode2020 process.

| Date | Contents | Person in charge |
|---|---|---|
| **2020-04-06** | ObsMode-SSG kick off telecon to start the ObsMode2020 process | ObsMode lead |
| **2020-04-16** | ObsMode2021 test plan review (Group A) | Tech. leads |
| **2020-04-17** | ObsMode2021 test plan review (Group B) | Tech. leads |
| **2020-04-21** | Test plan document submission | Tech. leads |
| **2020-04-21** | Test plan review process starts | Tech. leads and reviewers |
| | | |
| **2020-06-19** | Requirement deadline for ONLINE subsystem (for the 2020JUL RELEASE) | Tech. leads |
| | | |
| **2020-07-22** | ObsMode-SSG telecon: Mid-term progress review | Tech. leads |
| | | |
| **2020-07-27** | OT Phase I/II requirement discussion (capability bases) | OT and SSR SSSs and Tech. leads |
| **2020-07-28** | OT Phase I/II requirement discussion (capability bases) | OT and SSR SSSs and Tech. leads |
| | | |
| **2020-09-11** | Technical report submission for group A | Tech. leads |
| **2020-09-12** | Technical summary review process start (until Go/no-go telecon on Sep. 29) | Tech. leads and reviewers |
| **2020-09-14** | OT phase I/II requirement deadline (group A) for 2020NOV release | OT SSS and Tech. leads |
| **2020-09-29** | Go/no-go telecon for group A | All |
| **2020-10-02** | Final report submission for the group A capabilities | Tech. leads |
| | | |
| **2020-10-16** | Technical report submission for group B | Tech. leads |
| **2020-10-17** | Technical summary review process start (until Go/no-go telecon on Oct. 28) | Tech. leads and reviewers |
| **2020-10-19** | OT phase I/II requirement deadline (group B) for 2020DEC release | OT SSS and Tech. leads |
| **2020-10-28** | Go/no-go telecon for group B | All |
| **2020-10-30** | Final report submission for the group B capabilities | Tech. leads |
| | | |
| **2020-11-27** | Requirements submission to the OFFLINE system | Tech. leads |
| **2020-12-01** | Send data reduction method (Cycle 8) to ADAPT | ObsMode lead |
| | | |
| **2020-12-17** | ALMA Cycle 8 pre-announcement | DSO head |
| | | |
| **2020-12-31** | Cycle 8 THB and PG (ObsMode related) change submission deadline to the editors. | Tech. leads |
| | | |
| **2021-01-15** | ObsMode2020 summary report circulation to stakeholders | ObsMode lead |
| | | |

## 2.3. Documentation and regular communications

While the ObsMode process is led by JAO, global involvement and contributions are key. Therefore it is needed to interact with ALMA colleagues across teams physically-located in different places and working in different time zones. Facilitating regular communication is vital for the success of the Obsmode process. To this end, the ObsMode lead made all information available in written format and shared it in the project internal website (a confluence page). Furthermore, in order to create more verbal communication across the teams, a monthly ObsMode telecon (see Table 2) was organized in addition to the decision-making telecons. This frequent interaction triggered specific discussion items, such as ObsMode near/mid/long-terms priorities, which require experts from different areas within the projects (Commissioning experts, subsystem scientists, ICT staff, Science operation side experts). Finally, the review process was implemented systematically and on a capability basis for the first time in the ObsMode process. The test plan document and technical summary report were reviewed by external experts in order to support the commissioning teams's activities properly as well as to make the ObsMode process more visible to colleagues from various stakeholders (see Section 4).

**Table 2.** Summary of topics discussed in the ObsMode monthly telecon.

| Date | Discussion Topics | Participants |
|---|---|---|
| 2020-06-8 | APDM and ASDM change request<br>ObsMode inputs for Cycle 9 capabilities | ObsMode, SSGs, ICTs |
| 2020-07-13 | ObsMode2021 preparations | ObsMode |
| 2020-08-12 | Amplitude calibration<br>Optimization for the pointing observations | ObsMode and SSGs |
| 2020-09-09 | OT Phase I/II requirements for Cycle 8 | ObsMode |
| 2020-10-13 | Selfcalibration | ObsMode and SSGs |
| 2020-11-10 | ObsMode2021 priorities | ObsMode and SSGs |

## 2.4. Additional Considerations

### 2.4.1. Covid-19 Pandemic

Due to the covid-19 pandemic situation in 2020, ALMA decided to shut down the observatory in March, 2020. The shutdown significantly affected the ObsMode process since this hampered obtaining test data following the original timeline. An extraordinary "countdown" meeting was held on May 22, 2020 for pan-ALMA coordination. Following a meeting with the ALMA Board, the observatory decided to extend Cycle 7 one more year and delay the ObsMode timelines accordingly. From the ObsMode point of view, the originally-planned ObsMode2020 (before the covid-19 pandemic) was tied to Cycle 9 capability developments. After the covid-19 pandemic with revised timelines, ObsMode2020 was now tied to the new Cycle 8, aiming to offer some capabilities for PI science observations starting in October, 2021. In reality, we did not obtain any test data during the observatory shutdown in 2020. Hence the majority of the items from ObsMode2020 were carried over to ObsMode2021 (see Section 5). These will be tested in 2021 with the goal of offering them in Cycle 9 science observations, starting in October 2022 (see Figure 1).

**In this document, we use the term of "Cycle 8" for the newly-defined Cycle 8 after timeline revision, which will start in October 2021. Cycle 8 which was originally defined to be started in October 2020 (before the pandemic situation required a new timeline) will be stated as (old) Cycle 8 hereafter.**

### 2.4.2. Prioritization and resources

Overall ObsMode priorities were identified following the procedure described in section 2.1. However, limited availability of human resources as well as difficulty scheduling tests restricted making progress on specific capabilities. This therefore forced us to change the original priorities and timelines. In particular, two capabilities listed below had to change their timelines.

- **VLBI capabilities (GMVA and EHT):** Due to the pandemic this year, no coordinated observations, originally scheduled in April and October, 2020, had happened. Therefore the entire test plan had to be postponed to 2021.

- **Online software capability:** To develop this capability, direct support from developers is critical in addition to the ObsMode side of activity. We had to drop part of the plan due to the limited availability of human power from the developers group and this delayed the plan by one year.

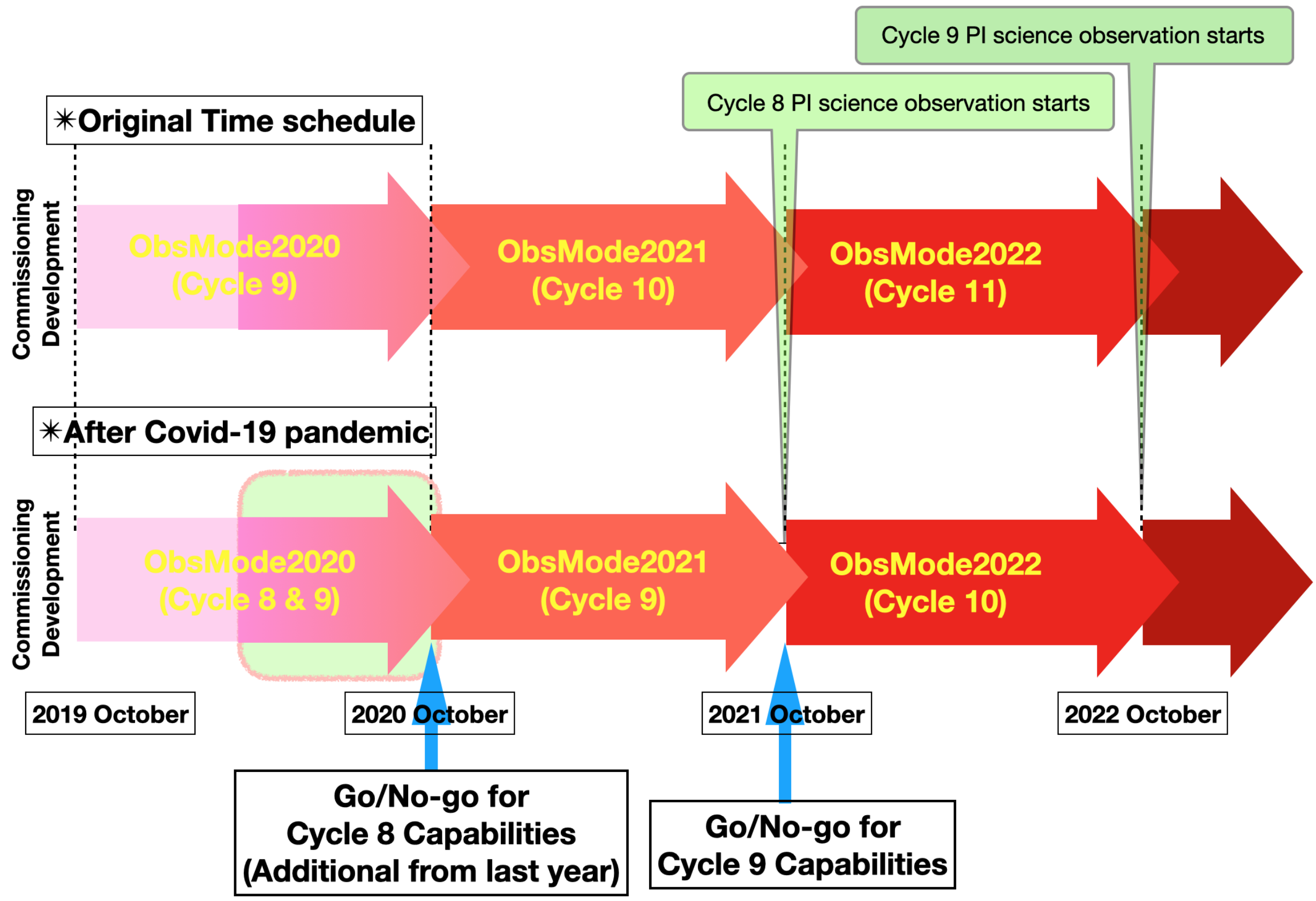

**Figure 1:** ObsMode2020 annual time schedule before (upper panel) and after (lower panel) the covid-19 pandemic re-scheduling.

#### 2.4.3. Phased array observing mode

The phased array observing mode is one of the capabilities that was already offered in 2019 as a part of the (old) Cycle 8 capability, mainly in order to observe pulsars. As with last year's preparations, software requirements and common software changes were discussed. However, the preparation process was running late with respect to the (old) Cycle 8 implementation timelines. In the end, the change was not ready for (old) Cycle 8. In the extraordinary meeting on May 22, 2020 (and in follow up discussion by ISOpT), it was officially decided to change the observing mode name from "Pulsar observing mode" to "Phased array mode". It was also decided that ObsMode2020 would take care of all changes required on the subsystem side, as well as in the common software, for using a new project code ".P". The policy for the observing mode, such as the scope of the science cases using this observing mode and the quality assurance procedure, were better determined (Section 5.1.2). Documentation of this observing mode was updated accordingly for Cycle 8, starting from October, 2021.

# 3. Roles and Responsibilities

In 2020, the ObsMode process was organized as an APG-led process. The head of APG at that moment, Eric Villard, prepared an initial framework for the process,including annual timelines and set priorities. In March 2020, Satoko Takahashi (a JAO operations astronomer) was assigned as the ObsMode2020 lead. Since then, she has been coordinating and running the entire ObsMode2020 process. The role of the ObsMode lead includes presenting the ObsMode2020 priorities, running the ObsMode review processes in order to assist the commissioning team in preparation of good test plans, and also supporting the commissioning team to make go/no-go decisions. Moreover, the ObsMode lead regularly communicates with the ObsMode technical leads (listed below) to track their work progress. She also communicates with subsystem scientists to make sure that the subsystem side of developments and preparations are also on track for the ObsMode2020-identified capabilities. In addition, the ObsMode lead also communicated with the head of APG (E. Villard) until July 2020, and the head of Science Operations (E. Humphreys) after August in order to make sure that the ObsMode efforts consistently align with the observatory-wide priorities, timelines, and activities.

Seven capabilities were identified as part of ObsMode2020. We identified technical leads for the seven capabilities as listed below. Roles of the technical leads are to create the commissioning team (small group of experts working on identified observing modes), lead commissioning test plan activities, prepare for the test plan documents, subsystem requirements, and technical summary reports to be ready for the capability go/no-go process. In addition, a few contacts (friends of some capabilities) are identified. They play the role of liaison between the external commissioning experts and JAO.

## 3.1. ObsMode2020 leads and contacts

- ObsMode lead
    - Satoko Takahashi (JAO operations astronomer)
        - Coordinator of the ObsMode2020 process

- Technical leads
    - Paulo Cortes (JAO operations astronomer)
        - Technical lead of the polarization capabilities
    - Tim Bastian (NRAO) and Masumi Shimojo (NAOJ)
        - Technical lead of the solar capabilities

- Satoko Takahashi (JAO operations astronomer)
  - Technical lead of the high frequency capabilities
- Yoshiharu Asaki (Commissioning scientist NAOJ, JAO)
  - Technical lead of the long baseline capabilities
- Ed Fomalont (Astronomer - Visitor program, NRAO, JAO)
  - Technical lead of the astrometry capabilities
- Geoff Crew (MIT Haystack Observatory)
  - Technical lead of the VLBI capabilities
- Lynn Matthews (MIT Haystack Observatory)
  - Technical lead of the VLBI capabilities
- Baltasar Vila-Vilaro (JAO system astronomer)
  - Technical lead of the online software capabilities

- Contact (friends of some capabilities)
  - Antonio Hales (JAO operations astronomer)
    - JAO contact point for the solar capabilities
  - Hugo Messias (JAO operations astronomer)
    - JAO contact point for the VLBI and phased array capabilities

# 4. Review Process

## 4.1. Procedure

For the ObsMode2020 process, the peer review process was newly implemented in order to receive extensive comments from technical experts within the project. Comments and feedback from the reviewers were provided for (i) the test plan documents and for (ii) the technical summary report.

The ObsMode test plan from each commissioning team was presented in the ObsMode-SSG telecons first in mid-April 2020, then the test plan documents from each team were submitted to the ObsMode lead on April 20, 2020. The test plan document review process was organized afterwards. Submitted test plan documents by the technical leads were circulated to the assigned ObsMode reviewers (Table 3). The reviewers had about a month to review the test plan document, and make first feedback to the commissioning team. Each commissioning team then spent time to discuss and finalize their test plan document. All the test plan documents are finalized by August, 2020 after a few iterations between the commissioning team and reviewers.

The technical summary reports submitted by the technical lead were also reviewed by the assigned reviewers. The reviewers had about two weeks to read and give feedback to the commissioning team before having the final capability go/no-go decision telecon in September and October, 2020.

The interactions between the commissioning team and reviewers were made via email, while telecon(s) could be set up to follow up some discussions, if necessary. All the inputs from the reviewers should be taken into account and properly addressed by the commissioning team before submitting the requested document as the final version.

The VLBI review process, including the selection of reviewers, requires approval from the AMT, which was coordinated with the deputy director (S. Corder). This is because the VLBI review process was formally requested by the ALMA board.

## 4.2. ObsMode2020 Reviewers

Reviewers selected for each capability are listed in Table 3. The reviewers were mostly selected by recommendations and suggestions from the technical leads. Additional suggestions were sometimes made by the ObsMode lead and head of APG to balance the specialties of the reviewers. The reviewers were technical experts of specific capabilities or those subsystem scientists who need to work closely with specific observing capabilities. Reviewers were mostly selected within the project (JAO or ARCs) and coordination was done between the ObsMode lead and their line managers. A few external reviewers were coordinated individually after approval from the head of DSO (E. Humphreys).

**Table 3:** ObsMode2020 reviewers

| Capability | Reviewer |
|---|---|
| **Baseline correlator** | Tsuyoshi Sawada (JAO), Alejandro Saez (JAO), and Dirk Petry (ESO) |
| **Solar (interferometer)** | Paulo Cortes (JAO) and Ed Fomalont (JAO) |
| **Solar (single dish)** | Richard Hills (Univ. of Cambridge) and Brian Mason (NRAO) |
| **Polarization** | Amanda Kepley (NRAO), Rosita Paladino (Italian ARC node), and George Moellenbrock (NRAO) |
| **High frequency** | Bill Dent (JAO), Baltasar Vila-Vilaro (JAO), and Neil Phillips (ESO) |
| **Long baseline** | Todd Hunter (NRAO), Catherine Vlahakis (NRAO), and Andy Biggs (ESO) |
| **Astrometry** | Ruediger Kneissl (JAO) and Leonid Petrov (NASA) |
| **VLBI** | Jim Braatz (NRAO), Seiji Kameno (JAO), Neil Phillips (ESO), Hiroshi Nagai (NAOJ), and Violette Impellizzeri (JAO→ Leiden ARC node) |

## 4.3. Reviewed documents

Reviewed documents are available on the ALMA internal web site. All the ALMA colleagues are able to access the information.

# 5. ObsMode2020 Go/no-go Review Process

In order to review ObsMode2020 progress and decide which tested new capabilities can be offered for Cycle 8, Go/No-go reviews and telecons were undertaken. The review process was performed dividing the capabilities into two groups: namely Group A and Group B. The Group A review process included Solar, Polarization, and Astrometry capabilities, while the Group B review process included Baseline correlator, VLBI, High frequency, and Long baseline. The review processes include technical report submissions (two weeks before the Go/no-go telecon). Review and iteration processes took place between the commissioning teams and reviewers, then the go/no-go review was held to conclude whether each capability would be offered in Cycle 8.

Due to the covid-19 pandemic in 2020, all the ObsMode2020 identified capabilities/observing modes were unable to obtain any test data on sky before the go/no-go review. Hence it is natural that most ObsMode2020 capabilities were carried over to the ObsMode2021 process. **It means that all the new capabilities except one described below are confirmed to be no-go for Cycle 8.**

The one capability that passed the go/no-go telecon, recommending it for an additional Cycle 8 capability, is **the 7m-array standalone full polarization observing mode (single-field limited).** In addition, preparations for the phased array mode (Section 2.4.3) and the definition of the current astrometry observing mode have been better clarified. These updates are available for observatory provided tools and user documentation.

In the following section, we describe progress and the current situation of each group. Note that these are only part of the capabilities/observing modes identified in the ObsMode2020 high priority items. No progress has been made for those not listed here due to the observatory shutdown (no commissioning activities have been made). Those prioritized items have been carried over to the ObsMode2021 process.

## 5.1. Capability-based summary

### 5.1.1. Polarization capabilities

The polarization commissioning team made progress on the following two items following the ObsMode2020 process.

- **7m-array full polarization capabilities (single-field, stand-alone mode)**

This is one of the highest priority items based on feedback from ASAC. The test data were obtained in 2019 in Band 3. The data were independently calibrated by two experts from the commissioning team (P. Cortes using CASA and S. Kameno using his own software), and a consistency check was made. In addition, ALMA calibrator surveys have been performed using full polarization mode and the database maintained by S. Kameno (AMAPOLA) has been used to select polarization calibrators for 12m-array PI science observations. Based on the test results and observatory experience, it was agreed that the 7m-array full polarization observing mode is technically ready for the single-pointing case. According to the decision, a OT requirement to activate this observing mode was accommodated in the Cycle 8 OT on time, and the capability was recommended to be offered for Cycle 8. Note that the mosaic test has not been completed yet, hence the mosaic capability will be carried over to the ObsMode2021 process.

ISOpT made the decision to support offering this capability at Cycle 8. However, in recognition of the work needed to process the data, ISOpT decided to consult with the DRMs to determine if there should be a cap placed on the number of hours offered to the community for this mode, and what that cap should be. This assessment was made within the DRM group (led by T. Nakos, the head of DMG) . On December 15, 2020, ISOpT accepted DMG's recommendation for a 75 hours cap (Cycle 8 decision).

- **Frequency setups for the mosaic linear polarization offered in (old) Cycle 8**

The 12m-array continuum mosaic observing mode with linear polarization capability was offered last year (Hull et al. 2020 PASP). While most of the testing was completed using TDM mode by the commissioning team, the observatory allowed PI to select FDM spectral setups (in order to allow for accurate removal of line contamination from the continuum emission) with arbitrary frequency. This could be a potential issue if a PI might select a frequency set up which is not optimized for continuum emission (or possibly observing line polarization, which is not offered in Cycle 8 yet). In ObsMode2020, we consulted experts within the project to determine the optimal frequency setups and the channel widths needed to reliably subtract the continuum emission from possible contamination by the line emission. After inputs, we agreed to use the single continuum setup with 1.8 GHz FDM windows in Cycle 8 (TDM mode is also available in the OT if the PI asks). This function is also updated in the Cycle 8 OT.

Whether arbitrary frequency setups can be allowed for the future depends on the progress of spectral line mosaic modes commissioning.

### 5.1.2. Phased array mode

The ALMA phased array mode allows us to phase-up all of the 12m antennas to function as a single telescope. This capability was already offered in (old) Cycle 8 in Band 3 and included in the call for proposals of March 2020. The phased array mode uses the same phasing technology that allows us to remove the requirement that VLBI targets have a flux density greater than 500 mJy.

**Software and subsystem regard:** Technical readiness was validated in the last two years by the VLBI commissioning team. However, subsystem and common software sides of the implementations were not ready yet and the policy side of the discussion since the last call needed advancement. During the ObsMode2020 process, we focused on improving these points. Integration into the common software was fully done in the 2020 December release (for Cycle 8). Most subsystems are now adopted to the phased array mode (using a new project code .p). The data archive system might need some work to ingest PSRFITS, which is the standard data format for the pulsar community. The work is in progress.

**QA2 and user documentation:** Last year (old Cycle 8), the project decided to not perform QA2 for the phased array mode. However, following discussion between the commissioning team and ISOpT on the potential impact of not performing QA2, and how the delivered data can be used for science, ISOpT decided following two points

- Cycle 8 phased array observing mode is only limited for observing known pulsars and also for pulsar searches. The mode will not be offered for non-pulsar science.
- At the time of writing this document, it is currently intended that ALMA will perform the light (quick) QA2 for the Cycle 8 phased array projects. Final decision will come from ISOpT.

Note that consulted stakeholders included the ALM A Deputy Director, the ALMA Friend of VLBI, the head of the DRM/DMG and the Obsmode Lead. The updated information will be updated on the user documentation before the Cycle 8 call for proposal will be released.

### 5.1.3. Astrometry

ALMA has not officially offered the astrometry observing mode (with extra calibrations such as better determining the antenna positions including atmospheric effect). There will be no change in Cycle 8. At the same time, the Cycle 8 OT has a scheme to auto-detect an "astrometry project", as indicated by users. The purpose is to auto-detect such projects instead of finding a key word of "astrometry" from >1800 proposals. This does not mean that we will guarantee the astrometric observing accuracy requested by the users, but the observatory will make best efforts to prepare the observations to have the best possible positional accuracies with the given conditions. The updates to the OT also align with planned future updates when an astrometric observing mode will be enabled, possibly in a few Cycles from now.

The technical lead (E. Fomalont) together with the OT Phase II group lead (H. Messias) collected recent data sets to reevaluate achievable astrometric accuracies with the phase referencing method currently used in ALMA science observations. This typically produces an astrometric accuracy of roughly 10% of FWHM beam size. Their analysis was also compared with those independently evaluated numbers by other experts, and the agreed number was updated in the Cycle 8 user documentation.

### 5.1.4. Solar capabilities

In Cycle 8, there will be no additional capabilities. The solar team has been working on expansion of solar observing capabilities such as the addition of polarimetry and improvement of the total power observations. These require major changes in the subsystems (some progress was made in 2020) and also a series of verification tests on sky (no recent progress due to the observatory shutdown in 2020).

### 5.1.5. Baseline correlator (Online related topics)

In Cycle 8, there are no updates for the baseline correlator projects, which are intended to improve the sensitivity and calibration accuracy of the scientific data sets. The test plans are carried over to the ObsMode2021 process.

.

### 5.1.6. High frequency observing capabilities

No progress was made on the commissioning due to the observatory shutdown in 2020, hence there will be no changes for the higher frequency observations (Band 8, 9, and 10) in the ACA 7m-array as well as the 12m-array observations with configurations C1 to C7 in Cycle 8. In ObsMode2021, the commissioning team will aim to enable the band-to-band observing mode, which observes the phase calibrator in lower frequency bands than that used for the science target. We determine the phase offset between the low- and high- frequency bands using the diff-gain -calibration source, then we apply the measured phase offset and scale with frequency to the phase calibrator and apply the calibration to obtain the high frequency science image. A significant advantage to observe the phase calibrator in the lower frequency band is to increase the availability of a closer calibrator to the science target. As an example, the typical angular distance ratio between finding the phase calibrator at high and low frequencies is about ~3 for Band 9 observations assuming the 2 GHz bandwidth. For Band 10, we do not always find a suitable phase calibrator for the ACA 7m-array. Commissioning the Band-to-band observing mode will be critical particularly for 7m-array observations in order to dramatically increase the number of high frequency science observations in the near future.

### 5.1.7. High frequency and Long Baseline

No progress was made on the commissioning due to the observatory shutdown in 2020. The commissioning team has been working on the B2B observation capability for high frequency (Band 8, 9, and 10) and long baseline configurations (Configs 8, 9, and 10, up to 16 km baselines) for observations to be ready for the near future Cycles. In order to make this capability ready, the commissioning team has been focusing on how to choose proper observing conditions as well as optimizing the observing parameters of the band-to-band method. Some of the initial test results obtained from previous years are available in published journals (Asaki et al. 2020 ApJS, Asaki et al. 2020 AJ, Maud et al. 2020 ApJS). From these previous studies, we more accurately determine the separation angle limitation and the switching cycle between the phase calibrator and target that will provide better quality of images. Further verifications and the subsystem side of implementations that are required to offer the high frequency long baseline capability will be tested in ObsMode2021.

# 6. Subsystem Readiness

The subsystem side of preparations and implementations are equally important as the technical readiness of each observing capability and observing mode. While no test data has been obtained due to the observatory shutdown, the time has been spent in discussing the subsystem side of preparations and implementations. Some of the most critical subsystems in terms of running ALMA observations are the OT (an offline subsystem) and the SSR (an online subsystem), which produce the scheduling block and make real time decisions (such as performing queries during observations) , respectively. These two subsystems communicate with each other. Without these two subsystems accepting the new observing capabilities and observing modes, even basic tests cannot be performed during the ObsMode process. This was the first year that all the ObsMode technical leads met with the subsystem scientists of the OT (A. Biggs) and the SSR (A. Hirota) and that requirements were coordinated among us. This resulted in not only implementing the necessary functions for Cycle 8 PI science observations, but also implementing necessary functions for future capabilities (i.e., capabilities prioritised for both ObsMode2020 and ObsMode2021). These new functions will be tested in the commissioning version of the OT in order to optimize the observing parameters. After a "go" decision for each capability, the finalized observing parameters will be migrated to the OT, which will be officially used for the corresponding ALMA Cycle.

# 7. User documentation

Based on the ObsMode2020 progress and Go/No-go telecon outcomes (Section 5), the technical handbook (THB) and proposers guide (PG) will be updated for Cycle 8. Items requiring updates (only ObsMode related topics) are collected by the ObsMode lead (Table 4).

After discussing with T. Remijan (former THB editor), P. Cortes (Cycle 8 THB editor) and J. Braatz (PG working group lead), we agreed on that only delta-changes will be performed for the Cycle 8 documentation because the Cycle 8 documentation work was already completed February 2020 (for old Cycle 8). Items listed in Table 4 are the minimum changes required based on work done during ObsMode2021. The changes are not only necessary from the technical side, but also include the policy side of decisions and clarifications. In order to make sure that the revisions are made appropriately, a few colleagues involved in each discussion played the role of a reviewer for the document revised by the technical leads. The finalized descriptions were delivered to the Cycle 8 THB editor (P. Cortes) and the PG working group lead (J. Braatz).

**Table 4:** PG and THB work assignments

| Topics | Authors | PG reviewers |
|---|---|---|
| **Astrometry** | Hugo Messias (JAO) and Ed Fomalont (JAO) | C. Vlahakis (NRAO), S. Martin (JAO), A. Biggs (ESO), and S. Takahashi (JAO) |
| **Linear polarization mosaic** | Paulo Cortes (JAO) | A. Biggs (ESO), H. Nagai (NAOJ), and S. Takahashi (JAO) |
| **7m-array polarization SACA** | Paulo Cortes (JAO) | H. Nagai (NAOJ), E. Humphreys (JAO), and T. Nakos (JAO) |
| **Phased array** | Hugo Messias (JAO), Lynn Matthews (MIT), and Geoff Crew (MIT) | M. Fukagawa (NAOJ), J. Braatz (NRAO), and S. Kameno (JAO) |
| **VLBI** | Hugo Messias (JAO), Lynn Matthews (MIT), and Geoff Crew (MIT) | H. Nagai (NAOJ), J. Braatz (NRAO), S. Kameno (JAO), and V. Impellizzeri (Leiden ARC node) |

# 8. Summary

The ObsMode2020 process took place between April to October, 2020.

- Seven capabilities were identified as high priority items for Cycle 9. Due to the observatory shutdown, the annual Cycle was delayed for one year. The ObsMode process also aligned with the change , hence the ObsMode2020 was adjusted to work for the Cycle 8preparations, however most of the identified capabilities have been carried over to ObsMode 2021, aiming to be offered for Cycle 9.

- The ObsMode review process has been implemented for the first time. A few (2-3) external reviewers were selected for each capability to review the test plan document and technical summary report provided by the technical leads. The reviewers also participated in the key ObsMode meetings and the go/no-go telecon to directly feedback their comments to the commissioning teams, in addition to email based iterations. The finalized reports are shared with colleagues in the ALMA project internal page as official ObsMode documents.

- The Cycle 8 go/no-go telecons were held on September 29 and October 28, 2020. The 7m-array polarization capability with ACA standalone mode (single- pointing) was recommended to be offered for Cycle 8 based on technical and subsystem readiness. Manual data processing will be performed for this mode, hence a 75-hour cap was recommended by the DRMs in order to not be overwhelmed with this task. ICT and the subsystem side performed preparations for accepting a new observing code, .P, for the phased array mode during ObsMode2020, and the mode will be ready for the Cycle 8. What can be provided for the astrometry observing mode was better defined and documented for Cycle 8. All the other high priority items had to be carried over to the ObsMode2021process (aiming to be ready for Cycle 9) due to the observatory shutdown in 2020. aiming to be ready for Cycle 9.

- The delta changes should be updated in the user documentation, which was coordinated with the editors of THB (Paulo Cortes) and PG (Jim Braatz).

- This ALMA memo is based on an ALMA EDM document (internal document: ALMA-90.00.00.00-0034-A-SPE) located at http://edm.alma.cl/forums/alma/dispatch.cgi/changereq/docProfileFol/103439

## 10. Acknowledgement

ObsMode process requires global involvement from all the executives including various stakeholders. The authors are grateful to all the ALMA colleagues who are directly and indirectly involved in the ObsMode2020 process. We particularly thank all the subsystem scientists for their support to the ObsMode activities. S. T. would like to thank Catherine Vlahakis (NRAO), Hiroshi Nagai (NAOJ), and Luke Maud (ESO), for their support to coordinate ObsMode activities with each ARC as a liaison. ST also thanks to John Carpenter (JAO) for providing us helpful comments and suggestions for this document.